\documentclass[10pt,journal,compsoc]{IEEEtran}

\usepackage{arydshln}
\usepackage{url}
\usepackage{amsfonts}
\usepackage{nicefrac}
\usepackage{amsmath}
\usepackage{dsfont}
\usepackage{amssymb}
\usepackage{graphicx}
\usepackage{tikz}
\usepackage[compress]{cite}

\newtheorem{theorem}{Theorem}[section]

\newtheorem{definition}{Definition}[section]
\newtheorem{proposition}[theorem]{Proposition}

\newtheorem{remark}{Remark}[section]

\newtheorem{assumption}{Assumption}[section]
\newtheorem{corollary}[theorem]{Corollary}

\newcommand{\st}{\mathop{\text{s.t.}}}
\newcommand{\sign}{\mathop{\text{sign}}}
\newcommand{\diag}{\mathop{\text{diag}}}
\newcommand{\argmin}{\mathop{\text{argmin}}}

\newcommand{\trace}{\mathrm{Tr}}

\newcommand{\supp}{\mathrm{supp}}

\newcommand{\citep}{\cite}

\newcommand{\revise}[1]{{\color{black}#1}}

\begin{document}

\title{Privacy-Preserving Public Release of Datasets for Support Vector Machine Classification }

\author{Farhad Farokhi,~\IEEEmembership{Senior Member,~IEEE}\vspace{-.1in}\IEEEcompsocitemizethanks{\IEEEcompsocthanksitem F.~Farokhi is with the CSIRO's Data61 and the University of Melbourne.\protect\\
e-mails: farhad.farokhi@unimelb.edu.au; farhad.farokhi@data61.csiro.au }}

\markboth{IEEE Transactions on Big Data,~Vol.~xx, No.~x, \today}%
{Farokhi: Privacy-Preserving Public Release of Datasets for Support Vector Machine Classification}

\IEEEtitleabstractindextext{
	\begin{abstract}
We consider the problem of publicly releasing a dataset  for support vector machine classification while not infringing on the privacy of data subjects (i.e., individuals whose private information is stored in the dataset). The dataset is systematically obfuscated using an additive noise for privacy protection. Motivated by the Cram\'{e}r-Rao bound, inverse of the trace of the Fisher information matrix is used as a measure of the privacy. Conditions are established for ensuring that the classifier extracted from the original dataset and the  obfuscated one are close to each other (capturing the utility). The optimal noise distribution is determined by maximizing a weighted sum of the measures of privacy and utility. The optimal privacy-preserving noise is proved to achieve local differential privacy. The results are generalized to a broader class of optimization-based supervised machine learning algorithms. Applicability of the methodology is demonstrated on multiple datasets.

\end{abstract}

}

\maketitle

\IEEEdisplaynontitleabstractindextext

\IEEEpeerreviewmaketitle

\IEEEraisesectionheading{\section{Introduction}}
\IEEEPARstart{A}{dvances}  in communication and information processing have opened new possibilities for data mining and big data analysis to answers  important challenges facing society. Public and private entities have thus scrambled to capitalize on these possibilities for improving the quality of offered services in a data-oriented manner. However, these enhancements come at the expense of the erosion of privacy within society. Therefore, there is a need for developing algorithms balancing utility and privacy.

Most often, governments would like to provide entire datasets to the public (or select private entities)\textit{ in a de-identified (anonymized) manner} so that academics, analysts, and researchers can utilize them for gaining valuable insights and developing new technologies.  Release of de-identified data can still infringe on the privacy of  people~\citep{sweeney2002k}. Therefore, methods, such as suppression and generalization, have been previously used  for private data release by guaranteeing $k$-anonymity. These methods also may not provide adequate individual privacy guarantees~\cite{1617392}.

This paper proposes a novel method for \textit{privacy-preserving release of an entire dataset while maintaining useful properties,
such as statistics required for reconstructing a support vector machine classifier}.  This is done by balancing privacy and utility guarantees using an explicit optimization problem. The dataset is systematically obfuscated using an additive noise and the inverse of the trace of the Fisher information matrix is used as a measure of privacy for the entries of the dataset. By the use of the Cram\'{e}r-Rao bound~\citep[p.\,169]{cramerraotheorem}, it can be seen that the defined measure of privacy provides a lower bound on the ability of an adversary estimating the individual entries of the dataset. The use of the Fisher information matrix makes the privacy metric independent of the sophistication of the adversary, thus making it a universal measure of privacy. Further, the Cram\'{e}r-Rao bound provides a practical/operational interpretation of the measure of privacy to the data owners, i.e., how much someone can learn about an individual in the dataset based on the publicly released obfuscated data. Conditions are provided for ensuring that the classifier extracted from the original data and the noisy data are ``close'' to each other. This serves as a measure of utility. To find the optimal noise distribution, a weighted sum of the measure of privacy and the measure of utility is maximized. In some cases, it is proved that \textit{the classifier extracted from the noisy privacy-preserving dataset is identical to the classifier extracted from the original private data.} Therefore, the utility is fully preserved while preserving the privacy. Finally, we show that the optimal additive noise from the presented framework also ensures $(\epsilon,\delta)$-local differential privacy. This is an interesting observation because (\textit{i}) the new methodology inherits the strong guarantees that come with the local differential privacy and (\textit{ii}) we can provide an operative interpretation for the parameters of $(\epsilon,\delta)$-local differential privacy  using the new measure of privacy in this paper and with the aid of the Cram\'{e}r-Rao bound.

In summary, this paper makes the following contributions in privacy-preserving release of datasets:\vspace{-.1in}
\begin{itemize}
	\item Using the Cram\'{e}r-Rao bound, an adversary's estimation error for reconstructing the individual entries of a private dataset  is related to the inverse of the trace of the Fisher information matrix. This enables the use of the Fisher information as a measure of privacy.
	\item The effect of additive privacy-preserving noise on the utility of machine learning models is captured. The utility of the machine learning models is negatively impacted by the variance of the additive noise. This is first proved for support vector machines and then generalized to optimization-based machine learning. 
	\item The optimal privacy-preserving noise distribution is computed by maximizing  a weighted sum of the measures of privacy and utility.  
	\item The optimal privacy-preserving is shown to be Gaussian and proved to satisfy $(\epsilon,\delta)$-local differential privacy. For additive privacy-preserving noises that can be correlated among the entries of the dataset, a special multi-dimensional Gaussian noise can be realized that does not degrade the utility of linear support vector machines while providing privacy. 
	\item An operative interpretation for the parameters of $(\epsilon,\delta)$-local differential privacy is provided by proving that the ability of an adversary estimating the individual entries of the dataset is bounded by $\mathcal{O}(\ln^2(\delta^{-1})\epsilon^{-2})$.  
	\item The effect of the optimal privacy-preserving noise is demonstrated on  the Wisconsin Breast Cancer Diagnostic dataset and the Adult dataset using linear support vector machines and the Lending Club using linear regression. The utility of the machine learning models trained on the data obfuscated by the optimal noise is always better than the utility of machine learning models trained on the data obfuscated by the Laplace mechanism for providing local differential privacy with the same privacy guarantee.
\end{itemize}

\subsection{Related Work}
%\subsubsection{Noiseless Dataset Release} 
Most often $k$-anonymity and its extensions, such as $\ell$-diversity and $m$-invariance, have been utilized for releasing datasets in a privacy-preserving manners~\cite{xiao2007m,zou2009k, machanavajjhala2006diversity}. These methods are however vulnerable to attacks; see, e.g.,~\cite{1617392,8662687,li2007t}. This demonstrate the need for the use of privacy-preserving noise when releasing datasets publicly. 

%\subsubsection{Differential Privacy} 
Differential privacy is an important methodology for privacy-preserving data release and handling because of possessing strong guarantees and post-processing properties~\citep{TCS-042}. Differential privacy has been successfully utilized in machine learning problems with applications to support vector machines~\cite{rubinstein2012learning,li2014privacy}, logistic regression~\cite{chaudhuri2009privacy,zhang2012functional}, and deep learning~\cite{abadi2016deep,shokri2015privacy}. These studies, however,  concentrate on responding to aggregate queries, such as releasing trained machine learning models in a privacy-preserving manner or reporting statistics of a dataset without jeopardising privacy of individuals. Another relevant notion of privacy is local differential privacy in which the data of individuals in perturbed \textit{locally} to ensure privacy, perhaps due to the lack \revise{of} trust in the aggregator. Although powerful in preserving privacy, differential privacy and local differential privacy often lack systematic approaches for selecting their parameters, also known as the privacy budget, in order to balance utility and privacy. This has caused concerns in effectiveness of implementations of differentially-private mechanisms in practice~\cite{tang2017privacy}. In this paper, with the aid of the Fisher information and the Cram\'{e}r-Rao bound, we also provide a systematic approach for setting the privacy budgets in local differential privacy when using the Gaussian mechanism.

%\subsubsection{Information-Theoretic Privacy}
An alternative to differential privacy is information-theoretic privacy, \revise{dating} back to wiretap channels~\cite{6772207} and their extension~\cite{yamamoto1983source}. Information-theoretic measures of privacy use mean square estimation error~\cite{farokhi2015quadratic}, mutual information~\cite{liang2009information}, and the
Fisher information~\cite{farokhisandberg2016} for measuring private information leakage. The use of entropy as a measure of privacy forces the private dataset to be statistically distributed with known distributions. For instance, it is assumed that the density of the private dataset is known~\cite{asoodeh2014notes} or that the data is Gaussian in~\cite{akyol2016information}. This weakens the results as the density of the data might not be available in advance or that data might follow the underlying assumptions. Therefore, in this paper, we use the Fisher information as a measure of private information leakage because it does not impose statistical assumptions on the private dataset.  In fact, when using the Fisher information, the private dataset is treated as an unknown arbitrary deterministic object with no assumptions about its origin. The Fisher information has been previously used for privacy preservation in~\cite{anderson1977efficiency,farokhisandberg2016,farokhi2019ensuring,farokhiSpringerChapter2020}. In those studies, however, the utility was not tailored to preserving quality of machine learning models as in this paper. For instance, in this paper, we show that a special multi-dimensional Gaussian noise can be used for privacy-preservation that does not degrade the utility of linear support vector machines. This would not have been possible following those studies.

Other approaches for privacy-preserving machine learning have been proposed previously~\cite{lin2008releasing}. These approaches also rely on privacy-preserving release of the machine learning model rather than the dataset. 

In image datasets, partial obfuscation of images to hide important private features, such as faces or license plates, have been considered~\cite{he2016puppies,olteanu2018consensual}. However, these studies do not easily generalize to other dataset. 

%\subsubsection{Synthetic Dataset}
The results of this paper are close, in spirit, to construction of synthetic datasets in which a dataset is generated to match statistics of the original dataset perturbed in a differentially-private manner; see, e.g.,~\cite{charest2011can,abowd2008protective}. Synthetic datasets are known to contain implausible entries (e.g., smoking infant) due to  only considering low order statistics (e.g., the first and second order statistics). This is because of the underlying computational complexity associated with realising synthetic datasets. Furthermore, generation of synthetic datasets requires stochasticity assumptions. 

\subsection{Organization}
The rest of the paper is organized as follows. The methodology is developed in Sections~\ref{sec:method}. Numerical results are demonstrated in Section~\ref{sec:numerical}. Finally, Section~\ref{sec:conclusions} concludes the paper and provides some directions for future research.

\section{Methodology} \label{sec:method}
We start with presenting the results for the simple case of linear support vector machines.  At the end of this section, in Subsection~\ref{sec:general}, we show that the results hold for more general optimization-based machine learning algorithms.
\subsection{Soft Margin Support Vector Machine}
Consider a set of training data $\{(x_i,y_i)\}_{i=1}^q\subseteq \mathbb{R}^p\times\{-1,+1\}$. In binary linear support vector machine classification, it is desired to obtain a separating hyper plane of the form $\{x\in\mathbb{R}^p: f(x)=\alpha^\top x+\beta=0\},$
with its corresponding classification rule
$\sign(f(x))$ to group the training data into two sets (of $y=+1$ and $y=-1$), i.e., ensure that $\sign(f(x_i))=y_i$ for all $i$. Up to re-scaling of $\alpha$ and $\beta$, this problem can be cast as
\begin{subequations}
\label{eqn:optim:0}
\begin{align}
\argmin_{\alpha\in\mathbb{R}^p,\beta\in\mathbb{R},\xi\in\mathbb{R}^q}  & \frac{1}{2} \alpha^\top \alpha+\theta\mathds{1}^\top \xi,
\\[-.6em]
\st \hspace{.3in} & y_i(\alpha^\top x_i+\beta)\geq 1-\xi_i, &&  \forall i\in\mathcal{Q},
\\
&\xi_i\geq 0, &&\forall i\in\mathcal{Q}.
\end{align}
\end{subequations}
where $\mathcal{Q}:=\{1,\dots,q\}$ and $\theta>0$. Note that $y_i(\alpha^\top x_i+\beta)>1$ implies that $y_i$ and $f(x)=\alpha^\top x_i+\beta$ have the same sign, i.e., the classifier correctly groups the training data $(x_i,y_i)$. Since it might not be possible to perfectly classify the data into two sets, there might be cases in which $\xi_i>1$ for some $i$. In those cases, the classifier miss-categorizes some data points. It is desired to keep the number of such entries as low as possible by adding the term $\theta\mathds{1}^\top \xi$ to the cost function. The optimization problem in~\eqref{eqn:optim:0} may not admit a unique solution. To alleviate this issue, an alternative problem can be posed:
\begin{subequations}
\label{eqn:optim:1}
\begin{align}
(\alpha^*,\beta^*,\xi^*):=\hspace{-.2in}\argmin_{\alpha\in\mathbb{R}^p,\beta\in\mathbb{R},\xi\in\mathbb{R}^q}  & \frac{1}{2} \alpha^\top \hspace{-.03in}\alpha\hspace{-.03in}+\hspace{-.03in}\frac{\rho}{2}(\beta^2\hspace{-.03in}+\hspace{-.03in}\xi^\top \xi)\hspace{-.03in}+\hspace{-.03in}\theta\mathds{1}^\top \xi,\\[-.6em]
\st \hspace{.3in} & y_i(\alpha^\top x_i\hspace{-.03in}+\hspace{-.03in}\beta)\hspace{-.03in}\geq\hspace{-.03in} 1\hspace{-.03in}-\hspace{-.03in}\xi_i, \forall i\hspace{-.03in}\in\hspace{-.03in}\mathcal{Q},
\\
&\xi_i\geq 0,\forall i\in\mathcal{Q},
\end{align}
\end{subequations}
where $\rho>0$. The uniqueness of the solution of~\eqref{eqn:optim:1} is guaranteed by the strict convexity of the cost and the convexity of the constraint set. Note that, by reducing $\rho$, the solution of the optimization problem in~\eqref{eqn:optim:1} can be made to arbitrarily closely approximate a solution of~\eqref{eqn:optim:0}. Therefore, the proposed alteration is without loss of generality; however, it simplifies the derivation of subsequent results.

\begin{proposition} \label{prop:rho} 
Let $\aleph$ denote the set of solutions of~\eqref{eqn:optim:0}. Then, $\lim_{\rho\searrow 0} (\alpha^*,\beta^*,\xi^*)\in\aleph$. 
\end{proposition}

\begin{IEEEproof} See Appendix~\ref{app:0}.
\end{IEEEproof}

\subsection{Independently and Identically Distributed Noise}
Solving~\eqref{eqn:optim:1}, or as a matter of fact~\eqref{eqn:optim:0}, requires access to the training data, which can infringe on the privacy of  individuals whose data is gathered for machine learning training. Therefore, the data owners are inclined to provide a noisy version of the data $\{(\bar{x}_i,y_i)\}_{i\in\mathcal{Q}}\subseteq \mathbb{R}^p\times\{-1,+1\}$ in which
\begin{align}\label{eqn:additive_local}
\bar{x}_i=x_i+n_i,
\end{align}
where $n_i\in\mathbb{R}^p$ is a zero-mean random additive noise with probability density function $\gamma:\mathbb{R}^p\rightarrow\mathbb{R}_{\geq 0}$. Note that $n_i$, $\forall i\in\mathcal{Q}$, are assumed to be drawn from the same distribution. Therefore, they are independently and identically distributed (i.i.d.) random variables. This makes process of generating the noise computationally friendly for very large databases (i.e., when $q$ is very large); however, as explained later, it also restricts the set of applicable noises. In the next subsection, the noise is generalized by removing the i.i.d. assumption. 

\begin{assumption} $\gamma$ is twice continuously differentiable.
\end{assumption}

Under this assumption, a lower-bound for an adversary's estimation error of the private database (based on the noisy data) can be determined. The lower bound is independent of the actions of the adversary and is thus immune to unrealistic assumptions (e.g., that the adversary is rational in case of game-theoretic approaches to privacy analysis). In what follows, $\supp(\gamma)$ is defined to be $\{n\,|\,\gamma(n)>0\}$. Further, for any continuously differentiable function $g(x)$, the notation $\partial g(x)/\partial x$ is reserved to denote a column vector containing the partial derivatives of the function\footnote{Note that there are other conventions in which the gradient is a row vector; however, in this paper, the gradient is a column vector.}. Further, for any twice continuously differentiable function $g(x)$, $\mathfrak{D}^2g(x)$ is the Hessian matrix (i.e., a symmetric matrix containing all the second order partial derivative). 

\begin{proposition} \label{prop:Fisher}
For any unbiased estimate of $x_i$ denoted by $\hat{x}_i$, it holds that
\begin{align} \label{eqn:cramer_rao}
\mathbb{E}\{\|\Pi(x_i-\hat{x}_i)\|_2^2\}
&\geq \trace(\Pi\mathcal{I}^{-1})\geq 1/\trace(\Pi^{-1}\mathcal{I}),
\end{align}
where $\Pi\in\mathbb{R}^{p\times p}$ is a positive definite matrix and $\mathcal{I}$ is the Fisher information matrix defined as
\begin{align}
\mathcal{I}
&=\int_{n\in \supp(\gamma)}\hspace{-.2in}
\gamma(n)\bigg[\frac{\partial \log(\gamma(n))}{\partial n} \bigg]\bigg[\frac{\partial \log(\gamma(n))}{\partial n} \bigg]^\top \mathrm{d}n.
\end{align}
\end{proposition}

\begin{IEEEproof}
The first inequality in~\eqref{eqn:cramer_rao} follows from the Cram\'{e}r-Rao bound~\citep[p.\,169]{cramerraotheorem}. The rest follows from \revise{the fact} that
$\trace(\Pi\mathcal{I}^{-1})
=\trace(\Pi^{1/2}\mathcal{I}^{-1}\Pi^{1/2})
\geq 1/\trace(\Pi^{-1/2}\mathcal{I}\Pi^{-1/2})
=1/\trace(\Pi^{-1}\mathcal{I})$
with the inequality stemming from Item~(11) in~\citep[p.\,44]{lutkepohl1997handbook}.
\end{IEEEproof}

Therefore, by minimizing $\trace(\Pi^{-1}\mathcal{I})$, the privacy of the participants can be guaranteed. This is because, irrespective of the adversaries behaviour (computational resources, intelligence, etc.), it cannot estimate the original entries of the database with an estimation error smaller than $1/\trace(\Pi^{-1}\mathcal{I})$. Note that the positive definite scaling matrix $\Pi$ is used here because not all entries of $x_i$ have the same scaling (e.g., some might be fractional numbers between zero and one while others are large integers). 

The addition of the noise potentially changes the classifier. Therefore, a measure of quality must be established (as otherwise the most private decision is to use an additive noise with co-variance approaching infinity). The optimal linear support vector machine can be extracted from
\begin{subequations}
\label{eqn:optim:2}
\begin{align}
(\bar{\alpha}^*,\bar{\beta}^*,\bar{\xi}^*):=\hspace{-.2in}\argmin_{\alpha\in\mathbb{R}^p,\beta\in\mathbb{R},\xi\in\mathbb{R}^q}  & \frac{1}{2} \alpha^\top\hspace{-.03in} \alpha\hspace{-.03in}+\hspace{-.03in}\frac{\rho}{2}(\beta^2\hspace{-.03in}+\hspace{-.03in}\xi^\top \xi)\hspace{-.03in}+\hspace{-.03in}\theta\mathds{1}^\top \xi,\\[-.6em]
\st \hspace{.3in} & y_i(\alpha^\top \bar{x}_i\hspace{-.03in}+\hspace{-.03in}\beta)\hspace{-.03in}\geq\hspace{-.03in} 1\hspace{-.03in}-\hspace{-.03in}\xi_i,\forall i\hspace{-.03in}\in\hspace{-.03in}\mathcal{Q},
\\
&\xi_i\geq 0,\forall i\in\mathcal{Q}.
\end{align}
\end{subequations}
Note that the difference between~\eqref{eqn:optim:1} and~\eqref{eqn:optim:2} is access to the noisy or original data. The solution of~\eqref{eqn:optim:2} is in fact a random variable. The quality of the classifier extracted from the noisy data can be guaranteed by ensuring that the following measure of quality is large:
\begin{align}
\mathcal{D}:=\mathbb{P}\left\{\left\|
\begin{bmatrix}
\bar{\alpha}^* -\alpha^*  \\ 
\bar{\beta}^* -\beta^*  \\
\bar{\xi}^* -\xi^* 
\end{bmatrix}
\right\|_2\leq \varepsilon\right\}.
\end{align}
The following result relates $\mathcal{D}$ to the second moment of the distribution of the additive noise.

\begin{proposition} \label{prop:probablity_lower_bound_quadratic} There exist $c\geq 1$ and $\varepsilon_0>0$ such that
$\mathcal{D}
\geq 1-(qc^2/\varepsilon^2)\trace(V_{nn}),\forall \varepsilon\in(0,\varepsilon_0)$ with $V_{nn}:=\mathbb{E}\{nn^\top\}$.
\end{proposition}

\begin{IEEEproof} See Appendix~\ref{app:1}.
\end{IEEEproof}

\begin{remark}
The proof of Proposition~\ref{prop:probablity_lower_bound_quadratic}, stemming from the application of \citep[Theorem~4.4]{daniel1973stability}, requires that the cost function of the classification optimization problem is positive definite. This is clearly the case for all choices of $\rho>0$. We can still select $\rho$ to be very small to reduce its impact on the classifier. This follows from Proposition 2.1 proving that the classifier with $\rho>0$ is close to the classifier with $\rho=0$ if $\rho$ is small enough. Therefore, restricting the classification problem to $\rho>0$ will not significantly or adversely influence the classifier if $\rho$ is small. Regarding privacy, we should note that the addition of Gaussian noise will always protect the privacy of the data; see Theorem~\ref{tho:del_eps} for guarantees in the sense of differential privacy. However, by selecting $\rho=0$, we cannot analyze the utility of privacy-preserving dataset. Therefore, we recommend selecting an infinitesimally small $\rho$ to also provide utility guarantees.
\end{remark}

Proposition~\ref{prop:probablity_lower_bound_quadratic} shows that, if the co-variance of the additive noise is kept small, $\mathcal{D}$ remains large. This motivates the following formulation for finding the optimal privacy-preserving noise:
\begin{align}  \label{eqn:optim_trace:eps_delta_dual}
\mathbf{P}(\lambda):\min_{\gamma}\; & \trace(\Pi^{-1}\mathcal{I})+\lambda \trace(V_{nn}).
\end{align}
We can also cast the problem of finding the optimal privacy-preserving noise as $\max_{\gamma}\; [1/\trace(\Pi^{-1}\mathcal{I})-\bar{\lambda} \trace(V_{nn})]$. However, these two problems are equivalent in the sense that, for any $\bar{\lambda}>0$, there exists $\lambda>0$ such that they admit the same solution. Now, one of the main results of the paper regarding the optimal privacy-preserving policy can be presented.

\begin{figure}[t]
\centering
\begin{tikzpicture}
\node[] at (0,0) {
\includegraphics[width=.39\linewidth]{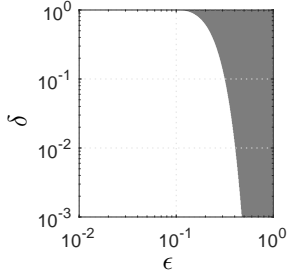}};
\node[] at (0,-1.9) {$\epsilon$};
\node[rotate=90] at (-2.0,0) {$\delta$};
\end{tikzpicture}
\vspace{-.15in}
\caption{\label{fig:eps_del_0} The gray region illustrates the set of $\delta$ and $\epsilon$ that meet the condition of Theorem~\ref{tho:del_eps} with $\sqrt{\lambda}\|\Pi^{1/2}\|_{\infty,2}=0.1$. }
\vspace{-.2in}
\end{figure}

\begin{figure}[t]
\centering
\begin{tikzpicture}
\node[] at (0,0) {
\includegraphics[width=0.49\linewidth]{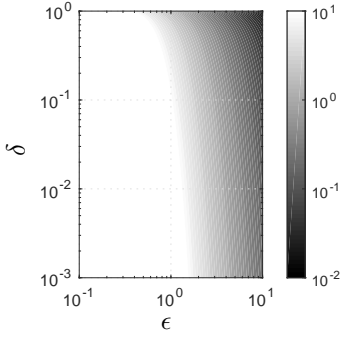}};
\node[] at (-.5,-2.3) {$\epsilon$};
\node[rotate=90] at (-2.5,0) {$\delta$};
\end{tikzpicture}
\vspace{-.15in}
\caption{\label{fig:eps_del_1} The lower bound on $\mathbb{E}\{\|\Pi(x_i-\hat{x}_i)\|_2^2\}$ for the case where $\trace(\Pi^{1/2})\|\Pi^{1/4}\|_{\infty,2}^2=1$.}
\vspace{-.2in}
\end{figure}

\begin{theorem} \label{tho:optimalnoise_1} The solution of $\mathbf{P}(\lambda)$ is given by 
\begin{align}\label{eqn:tho:condition:1}
\gamma^*(n)\hspace{-.03in}=\hspace{-.03in}\frac{1}{\sqrt{\det(2\pi \Pi^{-1/2}/\sqrt{\lambda})}}\exp\hspace{-.03in}\bigg(\hspace{-.05in}-\hspace{-.03in}\frac{\sqrt{\lambda}}{2}n^\top \Pi^{1/2} n\hspace{-.03in}\bigg).
\end{align}
\end{theorem}

\begin{IEEEproof} See Appendix~\ref{proof:tho:optimalnoise_1}.
\end{IEEEproof}

The optimal additive noise in Theorem~\ref{tho:optimalnoise_1}, which is a Gaussian noise, meets the condition of $(\epsilon,\delta)$-local differential privacy, presented below.

\begin{definition} The reporting policy in~\eqref{eqn:additive_local} is $(\epsilon,\delta)$-locally differential private if
$\mathbb{P}\{x_i+n_i\in\mathcal{X}\}\leq \exp(\epsilon)\mathbb{P}\{x'_i+n_i\in\mathcal{X}\}+\delta,$
where $\mathcal{X}\subseteq\mathbb{R}^p$ is a Borel-measurable set and $x_i,x'_i$ are possible entries in the dataset.
\end{definition}

\begin{theorem} \label{tho:del_eps} The optimal additive noise in Theorem~\ref{tho:optimalnoise_1} is $(\epsilon,\delta)$-locally differential private for all $\epsilon,\delta$ satisfying
$\lambda^{1/4}\|\Pi^{1/4}\|_{\infty,2}(1+\sqrt{2\ln(1/\delta)})\leq \epsilon,$
where $\|A\|_{\infty,2}=\max_{x\neq 0}\|Ax\|_2/\|x\|_\infty$ is the induced norm of $A$.
\end{theorem}

\begin{IEEEproof} The proof follows from combining the result of Theorem~\ref{tho:optimalnoise_1} and the differential privacy property of Gaussian additive noises in~\cite{nikolov2013geometry}.
\end{IEEEproof}

Theorem~\ref{tho:del_eps} proves that the optimal noise in  Theorem~\ref{tho:optimalnoise_1} is $(\epsilon,\delta)$-locally differential private. This is an important result as it shows that the new measure of privacy inherits the strong guarantees of differential privacy. The gray region in \revise{Fig.}~\ref{fig:eps_del_0} illustrates the set of $\delta$ and $\epsilon$ that meet the condition of Theorem~\ref{tho:del_eps} with $\lambda^{1/4}\|\Pi^{1/4}\|_{\infty,2}=0.1$. For $(\epsilon,\delta)$-locally differential private Gaussian mechanism, with the aid of Proposition~\ref{prop:Fisher_1}, we can see that 
$\mathbb{E}\{\|\Pi(x_i-\hat{x}_i)\|_2^2\}\geq \trace(\Pi^{1/2})\|\Pi^{1/4}\|_{\infty,2}^2(1+\sqrt{2\ln(1/\delta)})^2/\epsilon^2=\mathcal{O}(\ln^2(\delta^{-1})\epsilon^{-2}).
$
This inequality provides an operative iteration to the parameters of the differential privacy, i.e., we can bound on the error of estimating entries of the private database $\mathbb{E}\{\|\Pi(x_i-\hat{x}_i))\|_2^2\}$ for a specific selections of $(\epsilon,\delta)$. \revise{Fig.}~\ref{fig:eps_del_1} illustrates the lower bound on $\mathbb{E}\{\|\Pi(x_i-\hat{x}_i)\|_2^2\}$ for the case where $\trace(\Pi^{1/2})\|\Pi^{1/4}\|_{\infty,2}^2=1$. As expected, the adversary's estimation error grows  by decreasing $\epsilon$ and $\delta$.

In general, it might be desired to use noises whose support set is constrained to be a subset of $\mathcal{N}\subseteq\mathbb{R}^p$. This is because entries of database must be within some range to make sense (e.g., age or height cannot be negative). Note that, in this case, one cannot necessarily ensure that the additive noise has a zero mean (due to the arbitrary nature of the set $\mathcal{N}$). In this case, the problem can be cast as
\revise{
\begin{subequations}
\begin{align}  
\overline{\mathbf{P}}(\lambda):\min_{\gamma}\; & \trace(\Pi^{-1}\mathcal{I})+\lambda \trace(V_{nn}),\\
\st\;\, & \;\supp(\gamma)\subseteq \mathcal{N}.
\end{align}
\end{subequations}
}

\begin{theorem} The solution of $\overline{\mathbf{P}}(\lambda)$ is $\gamma^*(n)=\sqrt{\det(\Pi)}u(n)^2$, where
\begin{subequations}
\begin{align}
&\nabla^2 u(n)+(\mu-(\lambda/4) n^\top \Pi^{-1}n)u(n)=0,
\label{eqn:schrodinger}
\\[-.2em]
&u(n)=0, \forall n\in\partial \overline{\mathcal{N}},\quad u(n)>0,  \forall n\in\mathrm{int} (\overline{\mathcal{N}}),\\[-.2em]
&\int_{n\in\mathcal{N}}u(n)^2\mathrm{d}n=1,
\label{eqn:equality_prob}
\end{align}
for some $\mu>0$ with $\overline{\mathcal{N}}:=\{\Pi^{1/2}n,\forall n\in\mathcal{N}\}$.
\end{subequations}
\end{theorem}

\begin{IEEEproof} The proof follows from the same line of reasoning as in the proof of Theorem~\ref{tho:optimalnoise_1} with the change of variable $\gamma(n)=\sqrt{\det(\Pi)}u(\Pi^{1/2}n)^2$.
\end{IEEEproof}

Note that the partial differential equation~\eqref{eqn:schrodinger} is a stationary (time-independent) Schr\"{o}dinger equation. Therefore, it admits a unique solution for each $\mu$~\citep{mohamed2005separation}. Here, $\mu$ is the dual variable corresponding to the equality constraint in~\eqref{eqn:equality_prob} and can be computed using the dual ascent. 

\begin{corollary} \label{cor:1} Let $\Pi\hspace{-.03in}=\hspace{-.03in}\diag(\vartheta_1,\dots,\vartheta_p),\mathcal{N}\hspace{-.03in}=\hspace{-.03in}\prod_{i}[\underline{n}_i,\overline{n}_i]$. The solution of $\overline{\mathbf{P}}(\lambda)$ is $\gamma^*(n)=\sqrt{\prod_i \vartheta_i}(\prod_i u_i(n_i))^2$, where, for all $1\leq i\leq p$,
\begin{subequations}
\begin{align}
&\mathrm{d}^2 u_i(n_i)/\mathrm{d}n_i^2+(\mu_i-(\lambda/(4\vartheta_i)) n_i^2)u_i(n)=0,
\\
&u_i(n_i)\hspace{-.03in}=\hspace{-.03in}0, n_i\hspace{-.03in}\in\hspace{-.03in}\{\underline{n}_i,\overline{n}_i\}, u(n)>0, \forall n_i\in(\underline{n}_i,\overline{n}_i),\\
&\int_{\underline{n}_i}^{\overline{n}_i}u_i(n_i)^2\mathrm{d}n=1,
\end{align}
for some $\mu_i>0$.
\end{subequations}
\end{corollary}

\begin{IEEEproof} The proof follows from the method of separation of variables~\citep{mohamed2005separation}.
\end{IEEEproof}

The differential equations in Corollary~\ref{cor:1} are known as the Airy differential equations and their solution can be characterized using the Airy functions~\citep{airyref}. So far, the additive noise are somewhat  conservative as they are i.i.d. This assumption is removed in the next subsection.

\subsection{Correlated Noise}
In this section, it is assumed that the additive noise for various data points is correlated. Therefore, 
$
[
\bar{x}_1^\top \;
\cdots \;
\bar{x}_q^\top
]^\top
=
[
x_1^\top \;
\cdots \;
x_q^\top
]^\top
+w,
$
where $w\in\mathbb{R}^{qp}$ is a zero-mean random additive noise with probability density function $\tilde{\gamma}:\mathbb{R}^{qp}\rightarrow\mathbb{R}_{\geq 0}$. Similarly, the following standing assumption holds.

\begin{assumption} $\tilde{\gamma}(w)$ is twice continuously differentiable.
\end{assumption}

Again, the Cram\'{e}r-Rao bound can be used to establish a lower bound on the ability of the adversary for inferring the private database. In this paper, $\otimes$ and $\circ$, respectively, denote the Kronecker and the Hadamard products.

\begin{proposition} \label{prop:Fisher_1}
For any unbiased estimate of $x_i$ denoted by $\hat{x}_i$, it holds that
\begin{align}
\hspace{-.1in}\min_{1\leq i\leq q}\hspace{-.03in}\mathbb{E}\{\|\Pi(x_i-\hat{x}_i)\|_2^2\}
&\hspace{-.03in}\geq\hspace{-.03in} \min_{1\leq i\leq q}\hspace{-.03in}\trace(\Pi[(e_i^\top\hspace{-.03in} \otimes\hspace{-.03in} I_p)\widetilde{\mathcal{I}}(e_i\hspace{-.03in}\otimes\hspace{-.03in} I_p)]^{-1})\nonumber\\[-.5em]
&\geq 1/\trace((I_q\otimes \Pi^{-1})^{-1}\widetilde{\mathcal{I}}),
\end{align}
where $\widetilde{\mathcal{I}}$ is the Fisher information matrix defined as
\begin{align}
\widetilde{\mathcal{I}}
&=\int_{w\in \mathbb{R}^{qp}}\hspace{-.2in}
\tilde{\gamma}(w)\bigg[\frac{\partial \log(\tilde{\gamma}(w))}{\partial w} \bigg]\bigg[\frac{\partial \log(\tilde{\gamma}(w))}{\partial w} \bigg]^\top \mathrm{d}w.
\end{align}
\end{proposition}

\begin{IEEEproof} See Appendix~\ref{app:2}.
\end{IEEEproof}

The Karush--Kuhn--Tucker (KKT) conditions for~\eqref{eqn:optim:1} are \vspace{-.1in}
\begin{subequations}
\label{eqn:KKT}
\begin{align}
\alpha^* &=\sum_{i=1}^q (\omega^* )_iy_ix_i,
\quad
\beta^* =\sum_{i=1}^q (\omega^* )_iy_i/\rho,\\
(\xi^* )_i&=\max(((\omega^* )_i+(\varsigma^* )_i-1)/\rho,0), &&\hspace{-.15in}\forall i\in\mathcal{Q},\\
0&=(\omega^* )_i(y_i(\alpha^{*\top}  x_i+\beta^* )-1+(\xi^* )_i),&&\hspace{-.15in}\forall i\in\mathcal{Q},\\
0&=(\xi^* )_i(\varsigma^* )_i,&&\hspace{-.15in}\forall i \in\mathcal{Q},
\end{align}
\end{subequations}
where $\omega^*,\varsigma^*\in\mathbb{R}^q_{\geq 0}$ are Lagrange multipliers. The set of equations in~\eqref{eqn:KKT} have at least one solution due to the strict convexity of the cost and the convexity of the constraint set.

\begin{figure*}[t]
	\centering
	\begin{tabular}{ccc}
		\begin{tikzpicture}
		\node[] at (0,0) {
			\includegraphics[width=0.22\linewidth]{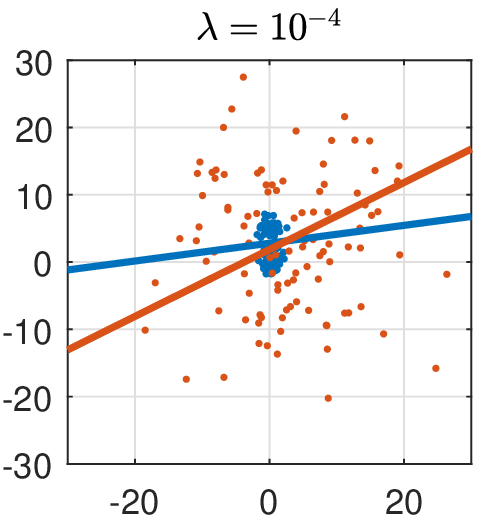}};
		\node[] at (0,-2.2) {$(x_i)_1$};
		\node[rotate=90] at (-2.2,0) {$(x_i)_2$};
		\end{tikzpicture}
		&
		\begin{tikzpicture}
		\node[] at (0,0) {
			\includegraphics[width=0.22\linewidth]{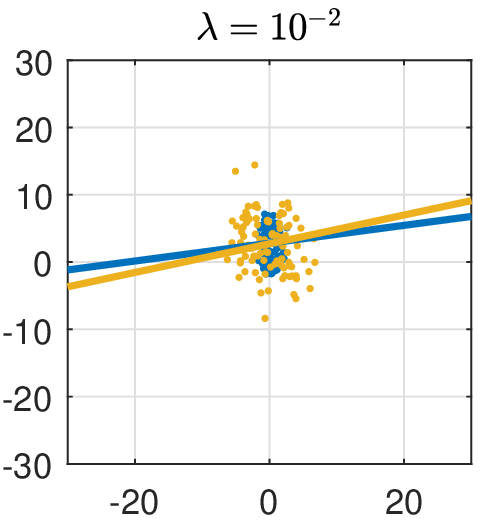}};
		\node[] at (0,-2.2) {$(x_i)_1$};
		\node[rotate=90] at (-2.2,0) {$(x_i)_2$};
		\end{tikzpicture}
		&
		\begin{tikzpicture}
		\node[] at (0,0) {
			\includegraphics[width=0.22\linewidth]{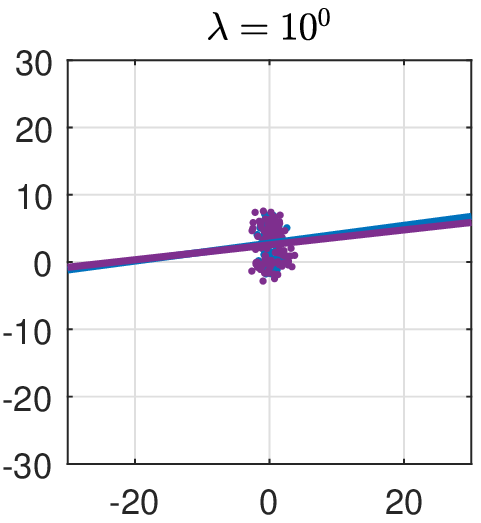}}; 
		\node[] at (0,-2.2) {$(x_i)_1$};
		\node[rotate=90] at (-2.2,0) {$(x_i)_2$};
		\end{tikzpicture}
	\end{tabular}
\vspace{-.15in}
	\caption{\label{fig:1} The blue dots illustrate the original data points $(x_i)_{i\in\mathcal{Q}}$ and the blue line is the classifier extracted from~\eqref{eqn:optim:1} using the original data. The red, yellow, and purple dots show the noisy data $(\bar{x}_i)_{i\in\mathcal{Q}}$ with the optimal noise in Theorem~\ref{tho:optimalnoise_1} for $\lambda=10^{-4}$, $\lambda=10^{-2}$, and $\lambda=10^{0}$, respectively. The \revise{solid} lines illustrate the classifiers extracted from~\eqref{eqn:optim:2} using the noisy data of the corresponding color. 
	}
\vspace{-.15in}
\end{figure*}

\begin{figure}
	\centering
	\vspace{-.2in}
	\begin{tikzpicture}
	\node[] at (0,0) {
		\includegraphics[width=0.46\linewidth]{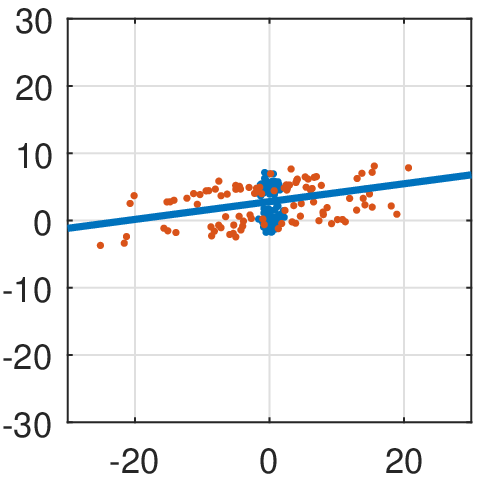}}; 
	\node[] at (0,-2.2) {$(x_i)_1$};
	\node[rotate=90] at (-2.2,0) {$(x_i)_2$};
	\end{tikzpicture}
	\vspace{-.2in}
	\caption{\label{fig:2} The blue dots illustrate the original data $(x_i)_{i\in\mathcal{Q}}$ and the blue line is the classifier extracted from~\eqref{eqn:optim:1} using the original data. The red dots illustrate the noisy data $(\bar{x}_i)_{i\in\mathcal{Q}}$ with the optimal noise in Theorem~\ref{tho:noisy_exact}. The classifier extracted from~\eqref{eqn:optim:2} using the noisy data remains the same. }
	\vspace{-.1in}
\end{figure}

\begin{proposition} \label{proposition:lastonetoremove}
$\mathbb{P}\hspace{-.03in}\left\{
\bar{\alpha}^*\hspace{-.03in}=\hspace{-.03in}\alpha^*,
\bar{\beta}^*\hspace{-.03in}=\hspace{-.03in}\beta^*,
\bar{\xi}^*\hspace{-.03in}=\hspace{-.03in}\xi^*
\hspace{-.03in}\right\}
\hspace{-.03in}= \hspace{-.03in}\mathbb{P}\hspace{-.03in}\left\{
\Omega
w
\hspace{-.03in}=\hspace{-.03in}0
\right\}$ with
$
\Omega:=[
((\lambda^{*}\circ y)^\top\otimes I_p)^\top\;
(I_q\otimes\alpha^{*\top} )^\top]^\top.
$
\end{proposition}

\begin{IEEEproof} See Appendix~\ref{proof:proposition:lastonetoremove}.
\end{IEEEproof}

This motivates the following problem formulation to find most privacy-preserving noise for which the classifiers extracted from the original data and the noisy data are identical:\begin{subequations} \label{eqn:optim:eps_delta}
\begin{align}
\widetilde{\mathbf{P}}:\min_{\tilde{\gamma}} \quad & \trace((I_q\otimes \Pi^{-1})\widetilde{\mathcal{I}}), \\[-.4em]
\st \hspace{.15in} &  V_{ww}\leq mI,
\\
&\mathbb{P}\left\{\Omega
w
=0\right\}=1,
\end{align}
\end{subequations}
where $V_{ww}$ is the co-variance of the additive noise, $m>0$, and $I$ is the identity matrix. The constraint $V_{ww}\leq mI$ is added to ensure that a noise with finite moment (thus realizable) is obtained. This is without any practical consequences as $m$ can be selected arbitrarily large. 

\begin{theorem} \label{tho:noisy_exact} Let $\Psi$ be a matrix whose columns form an orthonormal basis for the null space of $\Omega$. The solution of $\widetilde{\mathbf{P}}$ is 
$\tilde{\gamma}(w)=\mathds{1}_{w\in\mathrm{im}(\Psi)} \hat{\gamma}(\Psi^{\dag}w),$
where
$\hat{\gamma}(\bar{w})=((2\pi m)^{\mathrm{dim}(\mathrm{im}(\Psi))})^{-1/2}\exp(-\bar{w}^\top\bar{w}/(2m)).$
\end{theorem}

\begin{IEEEproof} See Appendix~\ref{proof:tho:noisy_exact}. 
\end{IEEEproof}

Theorem~\ref{tho:noisy_exact} provide\revise{s} the optimal noise for corrupting the dataset without any adverse effect on the utility of the linear support vector machine (i.e., no utility degradation).

\subsection{Optimization-Based Machine Learning}
\label{sec:general}
Here, we generalize the problem formulation of the previous subsections to more general optimization-based machine learning algorithms. Consider a set of training data $\{(x_i,y_i)\}_{i=1}^q\subseteq \mathbb{R}^{p_x}\times\mathbb{R}^{p_y}$. A general optimization-based machine learning problem can be cast as 
\begin{align} \label{eqn:general_optimization_ml}
\argmin_{\varphi\in\mathbb{R}^{p_\varphi}}\quad  & \ell(\varphi;\{(x_i,y_i)\}_{i=1}^q),
\end{align}
where $\ell$ is a fitness function. For instance, in nonlinear soft margin support vector machine
$
\ell(\varphi;\{(x_i,y_i)\}_{i=1}^q):=0.5 \varphi^\top \mathrm{diag}(I,0)
\varphi
+\theta\sum_{i=1}^q \max \left(1-y_i\begin{bmatrix}
\kappa(x_i)^\top & \hspace{-.1in}y_i
\end{bmatrix}\varphi,0\right),
$
where $\kappa(x_i)$ denotes the transformed data points using nonlinear mapping $\kappa(\cdot)$. Another example is the artificial neural networks with $\ell(\varphi;\{(x_i,y_i)\}_{i=1}^q): =\frac{1}{q}\sum_{i=1}^q \|y_i-\mathrm{ANN}(x_i,\varphi) \|_2^2,$ 
where $\mathrm{ANN}(x,\varphi)$ denotes the output of the artificial neural network.

\begin{proposition} \label{prop:lowerbound_general} Assume $\ell$ is twice continuously differentiable. There exist $c\geq 1$ and $\varepsilon_0\hspace{-.03in}>\hspace{-.03in}0$ such that $\mathcal{D}\hspace{-.03in}\geq\hspace{-.03in} 1\hspace{-.03in}-\hspace{-.03in}(q^2c^2/\varepsilon^2)\trace(V_{nn}),\forall \varepsilon\in(0,\varepsilon_0)$ with $V_{nn}\hspace{-.03in}:=\hspace{-.03in}\mathbb{E}\{nn^\top\hspace{-.03in}\}$.
\end{proposition}

\begin{IEEEproof} See Appendix~\ref{proof:prop:lowerbound_general}.
\end{IEEEproof}

The result of Proposition~\ref{prop:lowerbound_general} shows that the problem formulation~\eqref{eqn:optim_trace:eps_delta_dual} is still relevant for optimization-based machine learning. Therefore, Theorem~\ref{tho:optimalnoise_1} still provides the optimal additive noise when releasing datasets for more general machine learning algorithms.

\begin{remark}[Differentiability of Fitness Function] Proposition~\ref{prop:lowerbound_general} requires the fitness function to be twice differentiable. This is not the case generally. For instance, in deep learning, a popular non-differentiable activation function is rectified linear unit (ReLU). However, other activation functions, such as Sigmoid, softplus~\cite{dugas2001incorporating}, GELU~\cite{hendrycks2016gelu}, SoftExponential~\cite{godfrey2015continuum}, and SQNL~\cite{wuraola2018sqnl}, can be used to ensure twice-differentiability of fitness function. Again, note that the addition of a Gaussian noise will always protect the privacy of the data; see Theorem~\ref{tho:del_eps}. However, selecting a non-twice-differentiable loss function restricts our ability in analyzing the utility.
\end{remark}

\section{Numerical Example} \label{sec:numerical}
\subsection{Illustrative Example}
First, consider a simple example for the sake of the illustration of the results. A random dataset is generated with 50 entries drawn from a two-dimensional zero-mean Gaussian distribution with unit co-variance (corresponding to $y=+1$) and another 50 entries drawn from a two-dimensional Gaussian distribution with unit co-variance and mean $[0 \; 5]^\top$ (corresponding to $y=-1$). This dataset can be easily classified (visually), which is beneficial for the illustration of the effect of the noise. In what follows $\rho=10^{-2}$ so that the solutions of~\eqref{eqn:optim:0} and~\eqref{eqn:optim:1} are sufficiently close. Further, $\theta=1$ and $\Pi=I$.

First, the case with i.i.d. noise is considered. This is illustrated in \revise{Fig.}~\ref{fig:1}. The blue dots show the original data $(x_i)_{i\in\mathcal{Q}}$ and the blue line is the classifier extracted from~\eqref{eqn:optim:1} using the original data. The red, yellow, and purple dots show the noisy data $(\bar{x}_i)_{i\in\mathcal{Q}}$ with the optimal noise in Theorem~\ref{tho:optimalnoise_1} for $\lambda=10^{-4}$, $\lambda=10^{-2}$, and $\lambda=10^{0}$, respectively. The solid lines illustrate the classifiers extracted from~\eqref{eqn:optim:2} using the noisy data of the corresponding color. As expected, the classifiers constructed using the noisy data approaches the classifier extracted from the original data as $\lambda$ increases. Note that $\mathbb{E}\{\|x_i-\hat{x}_i\|_2^2\}$ is lower bounded by $200$, $20$, and $2$ for $\lambda=10^{-4}$, $\lambda=10^{-2}$, and $\lambda=10^{0}$, respectively.

\begin{figure}
\centering
\begin{tabular}{c}
	Breast Cancer Wisconsin (Diagnostic) dataset
	\\[-.7em]
\begin{tikzpicture}
\node[] at (0,0) {
\includegraphics[width=0.46\linewidth]{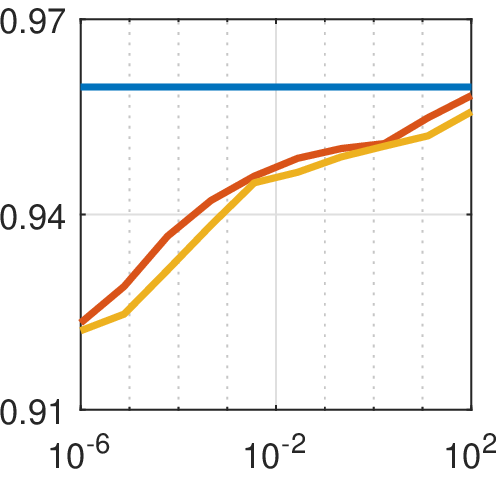}}; 
\node[] at (0,-2.2) {$\lambda$};
\node[rotate=90] at (-2.4,0) {\footnotesize $\mathbb{E}\{\sum \mathds{1}_{y_i(\bar{\alpha}^{*\top}x_i+\bar{\beta}^{*})>0}\}/n$};
\end{tikzpicture}
\\[-1.1em]
\begin{tikzpicture}
\node[] at (0,0) {
\includegraphics[width=0.46\linewidth]{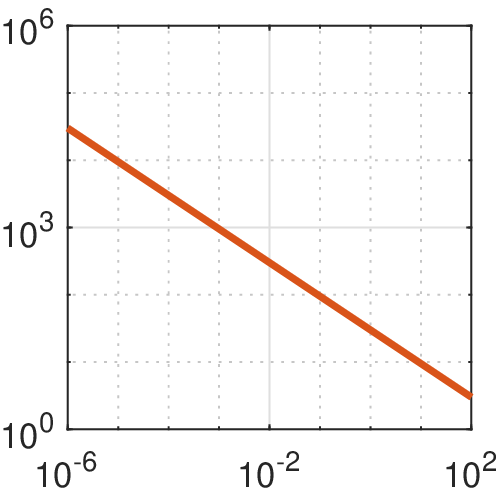}}; 
\node[] at (0,-2.2) {$\lambda$};
\node[rotate=90] at (-2.4,0) {\footnotesize $\min_{i}\mathbb{E}\{\|x_i-\hat{x}_i(\bar{x})\|_2^2\}$};
\end{tikzpicture}
\end{tabular}
\vspace{-.1in}
\caption{\label{fig:3} (top) The blue line illustrates the classification success rate of the classifier generated from the original data. The red curve depicts the  classification success rate using the classifier extracted from~\eqref{eqn:optim:2} using the noisy data corrupted with the optimal noise in Theorem~\ref{tho:optimalnoise_1} versus $\lambda$. The yellow curve shows the classification success rate based on the dataset obfuscated using Laplace noise to achieve local differential privacy (with the same privacy guarantee). (bottom) The privacy guarantee $\min_{i}\mathbb{E}\{\|x_i-\hat{x}_i\|_2^2\}$ versus~$\lambda$. }
\vspace{-.2in}
\end{figure}

Now, the case with correlated additive noise is considered. \revise{Fig.}~\ref{fig:2} shows the results in this case. The blue dots illustrate the original data $(x_i)_{i\in\mathcal{Q}}$ and the blue line is the classifier extracted from~\eqref{eqn:optim:1} using the original data. The red dots illustrate the noisy data $(\bar{x}_i)_{i\in\mathcal{Q}}$ with the optimal noise in Theorem~\ref{tho:noisy_exact} for $m=100$. Note that using this noise, the extracted classifier from~\eqref{eqn:optim:2} is the same as the classifier extracted from the original data. In this case, $\mathbb{E}\{\|x_i-\hat{x}_i\|\}\geq 80.5$. This error can be worsened by increasing $m$. The specific structure of the noise spreads the data along the classifier (and not towards it).

\subsection{Experimental Results}
Here, the applicability of the presented methodology on practical datasets is demonstrated with \revise{two machine learning algorithms. Particularly, we use the Breast Cancer Wisconsin (Diagnostic) dataset and the Adult dataset with linear support vector machines and the Lending Club dataset with linear regression. }
\subsubsection{Medical Dataset}
\revise{The Breast Cancer Wisconsin (Diagnostic) Data Set is openly available in~\citep{Dua:2017} containing features computed from a digitized image of a fine needle aspirate of breast mass. The features include parameters, such as radius, texture, smoothness, and symmetry, of cell nucleus.  For this dataset, we are interested in training a support vector machine for cancer diagnosis. We systematically corrupt the entries of the dataset to avoid the unintentional private data leakages. }

First, the case of i.i.d. additive noise is demonstrated. The red curve in \revise{Fig.}~\ref{fig:3}~(top) shows $\mathbb{E}\{\sum_{i=1}^q \mathds{1}_{y_i(\bar{\alpha}^{*\top}x_i+\bar{\beta}^{*})>0}\}/n$, empirically computed by averaging 100 simulations, versus $\lambda$. In this case, the classifier is extracted from the noisy data with the optimal noise in Theorem~\ref{tho:optimalnoise_1}. Note that $\mathbb{E}\{\sum_{i=1}^q \mathds{1}_{y_i(\bar{\alpha}^{*\top}x_i+\bar{\beta}^{*})>0}\}/n$ is the classification success rate of the original data using the classifier generated from the noisy data. The blue line illustrates the success rate of the classifier extracted from the original data.  As $\lambda$ increases, the success rate of the classifier from the noisy data approaches the classification rate of the classifier based on the original data. The yellow curve shows the classification rate for the case where the dataset is corrupted using Laplace noise to achieve  local differential privacy. The level of the differential privacy is set so that the the privacy guarantee $\min_i\mathbb{E}\{\|x_i-\hat{x}_i\|\}$ remains the same for both noises. This is to ensure that methods are compared in a fair manner. When using the Laplace mechanism, the success rate of the classifier is always below the optimal noise in Theorem~\ref{tho:optimalnoise_1}. The red curve in \revise{Fig.}~\ref{fig:3}~(bottom) illustrates the level of privacy $\min_i\mathbb{E}\{\|x_i-\hat{x}_i(\bar{x})\|\}$ versus $\lambda$. \revise{Fig.}~\ref{fig:3} captures the trade-off between privacy and utility.

The correlated additive noise can also be utilized in this example. Upon using the noisy data $(\bar{x}_i)_{i\in\mathcal{Q}}$ with the optimal noise in Theorem~\ref{tho:noisy_exact}, $\min_{i\in\mathcal{Q}}\mathbb{E}\{\|x_i-\hat{x}_i\|\}\geq 28.4m$. Therefore, for $m=100$, $\min_{i\in\mathcal{Q}}\mathbb{E}\{\|x_i-\hat{x}_i\|\}\geq 2840$. Note that, in this case, the classifier remains exactly the same even with such a high privacy guarantee.

\begin{figure}
	\centering
	\begin{tabular}{c}
		Adult dataset \\[-.7em]
		\begin{tikzpicture}
		\node[] at (0,0) {
			\includegraphics[width=0.46\linewidth]{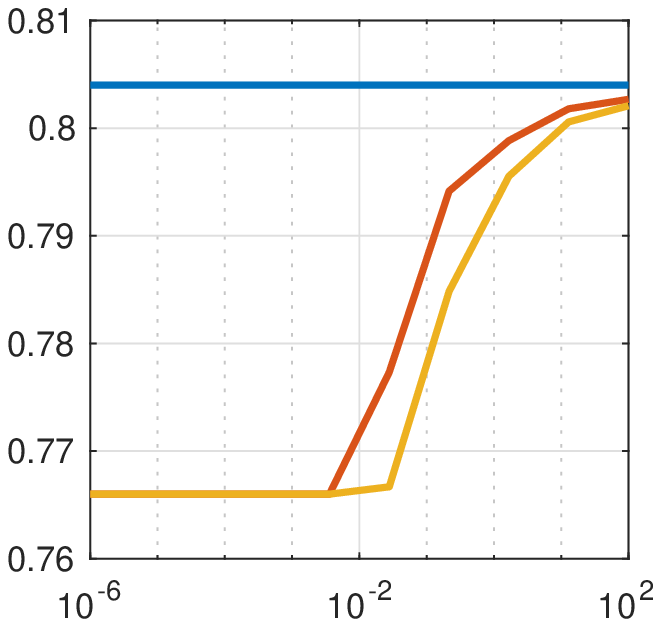}}; 
		\node[] at (0,-2.2) {$\lambda$};
		\node[rotate=90] at (-2.3,0) {$\mathbb{E}\{\sum \mathds{1}_{y_i(\bar{\alpha}^{*\top}x_i+\bar{\beta}^{*})>0}\}/n$};
		\end{tikzpicture}
	\end{tabular}
	\vspace{-.1in}
	\caption{\label{fig:3_5}  The blue line illustrates the classification success rate of the classifier generated from the original data. The red curve depicts the  classification success rate using the classifier extracted from~\eqref{eqn:optim:2} using the noisy data corrupted with the optimal noise in Theorem~\ref{tho:optimalnoise_1} versus $\lambda$. The yellow curve shows the classification success rate based on the dataset obfuscated using Laplace noise to achieve local differential privacy (with the same privacy guarantee). }
	\vspace{-.1in}
\end{figure}

\begin{figure}
\centering
\begin{tabular}{c}
	Lending Club dataset 
	\\[-.4em]
\begin{tikzpicture}
\node[] at (0,0) {
\includegraphics[width=0.46\linewidth]{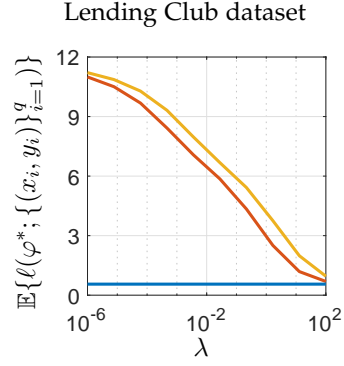}}; 
\node[] at (0,-2.2) {$\lambda$};
\node[rotate=90] at (-2.3,0) {$\mathbb{E}\{\ell(\varphi^*;\{(x_i,y_i)\}_{i=1}^q)\}$};
\end{tikzpicture}
\end{tabular}
\vspace{-.1in}
\caption{\label{fig:4} The blue line illustrates the fitness of regression model  generated from the original data. The red curve depict the fitness of the regression model extracted from~\eqref{eqn:general_optimization_ml} using the noisy data corrupted with the optimal noise in Theorem~\ref{tho:optimalnoise_1} versus $\lambda$. The yellow curve shows the fitness of regression model based on the dataset obfuscated using Laplace noise to achieve local differential privacy (with the same privacy guarantee). }
\vspace{-.1in}
\end{figure}

\subsubsection{Adult Dataset} 
\revise{This dataset contains nearly 49,000 records from the 1994 Census database~\cite{Dua:2017}. The records contain features, such as age, education, and work. The aim is to train a classifier that identifies individuals earning more than \$50,000. Here, we obfuscate the dataset in order to eliminate private information leakages about the individuals in the dataset.

We illustrate the effect of the optimal privacy-preserving noise in Theorem~\ref{tho:optimalnoise_1} in Fig.~\ref{fig:3_5}. The red curve in \revise{Fig.}~\ref{fig:3} shows $\mathbb{E}\{\sum_{i=1}^q \mathds{1}_{y_i(\bar{\alpha}^{*\top}x_i+\bar{\beta}^{*})>0}\}/n$, empirically computed by averaging 100 simulations, versus $\lambda$. The blue line illustrates the success rate of the classifier extracted from the original data. The success rate of the classifier from the noisy data approaches the classification rate of the classifier based on the original data as $\lambda$ increases. The yellow curve shows the classification success rate for the case where the dataset is corrupted using Laplace noise to achieve  local differential privacy. Again, the level of the differential privacy is set so that the the privacy guarantee, $\min_i\mathbb{E}\{\|x_i-\hat{x}_i\|\}$, remains the same for both noises. Evidently, the  optimal privacy-preserving noise in Theorem~\ref{tho:optimalnoise_1} outperforms the  Laplace noise in terms of the utility of the trained support vector machine model.
}

\subsubsection{Finance Dataset} \revise{This dataset contains information about approximately 890,000 loans made on the Lending Club, which is a peer-to-peer lending platform~\cite{kaggle1}. The dataset contains information about the loans, such as total amount, information about the borrower, such as age and credit rating, and the interest rates of the loans. We encode categorical data, such as state of residence and loan grade, with integer numbers. We also remove unique identifier attributes, such as member id, and unrelated attributes, such as the URL for the listing data. We are interested in training a linear regression $\varphi^\top x_i$ to estimate the loan interest rate $y_i$. Training the regression model can be cast as~\eqref{eqn:general_optimization_ml} with loss function $\ell(\varphi;\{(x_i,y_i)\}_{i=1}^q):=\sigma \varphi^\top \varphi+\sum_{i=1}^q (y_i-\varphi^\top x_i)^2/q$ for $\sigma>0$. In what follows, we set $\sigma=10^{-5}$. 

\revise{Fig.}~\ref{fig:4} shows the expected loss  $\mathbb{E}\{\ell(\varphi^*;\{(x_i,y_i)\}_{i=1}^q)\}$, empirically computed by averaging 100 simulations, as a function of $\lambda$. The red curve captures the fitness of the regression model generated from the noisy data with the optimal noise in Theorem~\ref{tho:optimalnoise_1}. The blue line illustrates the fitness of the regression model extracted from the original data without privacy preserving noise.  As $\lambda$, or the emphasis on utility, increases the fitness of the regression model from the noisy data approaches the fitness of the regression model computed based on the original data. The yellow curve shows the fitness rate for the case where the dataset is corrupted using Laplace noise to achieve local differential privacy. The fitness of the regression model with the optimal noise in Theorem~\ref{tho:optimalnoise_1}  is always better than the one with Laplace noise. }

\section{Conclusions and Future Work} \label{sec:conclusions}
A private dataset was corrupted by an additive noise for privacy protection in such a way that the classifier extracted from it remains close to the classifier computed based on the original dataset. The applicability of the methodology was demonstrated on  practical datasets in medicine, demography, and finance. Future work can focus on implementation of constrained additive noises for practical datasets.

%\subsubsection*{Acknowledgments}

\bibliographystyle{ieeetr}
\bibliography{ref}

\newpage

\appendices

\section{Proof of Proposition~\ref{prop:rho}}
\label{app:0}
Define the set valued mapping $S_{\mathrm{opt}}:\mathbb{R}\rightrightarrows\mathbb{R}^p\times\mathbb{R}\times \mathbb{R}^q$ such that 
\begin{align*}
S_{\mathrm{opt}}(p):=\bigg\{&(\alpha',\beta',\xi')\in S_{\mathrm{feas}}\,\bigg|\,\forall (\alpha,\beta,\xi)\in S_{\mathrm{feas}}:\\&\frac{1}{2} \alpha^{\prime\top} \alpha'+\frac{\rho}{2}(\beta^{\prime 2}+\xi^{\prime\top} \xi')+\theta\mathds{1}^\top \xi'\\
&
\quad \leq \frac{1}{2} \alpha^\top \alpha+\frac{\rho}{2}(\beta^2+\xi^\top \xi)+\theta\mathds{1}^\top \xi \bigg\},
\end{align*}
where
\begin{align*}
S_{\mathrm{feas}}:=\bigg\{&(\alpha,\beta,\xi)\in\mathbb{R}^p\times\mathbb{R}\times \mathbb{R}^q\,\bigg|\,y_i(\alpha^\top x_i+\beta)\geq 1-\xi_i, \\ &\forall i\in\mathcal{Q}\; \wedge\; \xi_i\geq 0, \forall i\in\mathcal{Q}\bigg\}.
\end{align*}
By definition, $S_{\mathrm{opt}}(p)$ is the set of the solutions of~\eqref{eqn:optim:1} for all $p$ (including $p=0$). Therefore, for all $p>0$, $S_{\mathrm{opt}}(p)=\{(\alpha^*,\beta^*,\xi^*)\}$ and $S_{\mathrm{opt}}(0)=\aleph$. From Theorem~3B.5 in~\citep{dontchev2014implicit}, it can be deduced that $\limsup_{p\rightarrow 0} S_{\mathrm{opt}}(p)\subseteq S_{\mathrm{opt}}(0)=\aleph.$ The rest follows from \revise{the fact} that 
$\lim_{\rho\searrow 0}(\alpha^*,\beta^*,\xi^*)=\lim_{\rho\searrow 0}S_{\mathrm{opt}}(p)\subseteq \limsup_{\rho\rightarrow 0}S_{\mathrm{opt}}(p)\subseteq\aleph$ where the first equality follows from \revise{the fact} that $S_{\mathrm{opt}}(p)$ is a singleton for all $p>0$. 

\section{Proof of Proposition~\ref{prop:probablity_lower_bound_quadratic}}
\label{app:1}
First, it is established that there exist $c\geq 1$ and $\epsilon_0>0$ such that 
\begin{align*}
\left\|
\begin{bmatrix}
y_1(x_1-\bar{x}_1)^\top \\ 
\vdots \\
y_q(x_q-\bar{x}_q)^\top
\end{bmatrix}
\right\|_2\leq \epsilon<\epsilon_0
\implies
\left\|
\begin{bmatrix}
\bar{\alpha}^* -\alpha^*  \\ 
\bar{\beta}^* -\beta^* \\
\bar{\xi}^* -\xi^* 
\end{bmatrix}
\right\|_2\leq c\epsilon.
\end{align*}
The proof of this claim follows from~\citep[Theorem~4.4]{daniel1973stability} and reorganizing the inequality constraints in~\eqref{eqn:optim:1} and~\eqref{eqn:optim:2} to be of the form of
\begin{align*}
-\begin{bmatrix}
y_1x_1^\top & y_1 \\ 
\vdots & \vdots \\
y_qx_q^\top & y_q
\end{bmatrix}
\begin{bmatrix}
\alpha\\ 
\beta
\end{bmatrix}
\leq -1 \mathds{1},\quad 
-\begin{bmatrix}
y_1\bar{x}_1^\top & y_1 \\ 
\vdots & \vdots \\
y_q\bar{x}_n^\top & y_q
\end{bmatrix}
\begin{bmatrix}
\alpha\\ 
\beta
\end{bmatrix}
\leq -1 \mathds{1}.
\end{align*}
Note that
\begin{align*}
\left\|
\begin{bmatrix}
y_1(x_1-\bar{x}_1)^\top \\ 
\vdots \\
y_n(x_q-\bar{x}_q)^\top
\end{bmatrix}
\right\|_2
&=\max_{\|p\|_2=1} \left\|
\begin{bmatrix}
y_1(x_1-\bar{x}_1)^\top \\ 
\vdots \\
y_n(x_q-\bar{x}_q)^\top
\end{bmatrix} p
\right\|_2\\
&=\max_{\|p\|_2=1} \left\|\begin{bmatrix}
y_1n_1^\top \\ 
\vdots \\
y_qn_q^\top
\end{bmatrix} p\right\|_2\\
&\leq \sqrt{\max_{\|p\|=1}\sum_{i=1}^q (y_in_i^\top p)^2}\\
&=\sqrt{\max_{\|p\|=1}\sum_{i=1}^q (n_i^\top p)^2}\\
&\leq \hspace{-.03in}\sqrt{\max_{\|p\|=1}\sum_{i=1}^q \|n_i\|_2^2 \|p\|_2^2}\hspace{-.03in}=\hspace{-.03in}\sqrt{\sum_{i=1}^q \|n_i\|_2^2}.
\end{align*}
This implies that
\begin{align*}
\mathbb{P} \left\{\left\|
\begin{bmatrix}
y_1(x_1-\bar{x}_1)^\top \\ 
\vdots \\
y_q(x_q-\bar{x}_q)^\top
\end{bmatrix}
\right\|_2\leq \epsilon \right\}
&\geq \mathbb{P}\left\{\sqrt{\sum_{i=1}^q \|n_i\|_2^2}\leq \epsilon\right\}\\
&= \mathbb{P}\left\{\sum_{i=1}^q \|n_i\|_2^2\leq \epsilon^2\right\}
\end{align*}
Therefore, it can be proved that
\begin{align*}
\mathbb{P}\left\{\left\|
\begin{bmatrix}
\bar{\alpha}^* -\alpha^*  \\ 
\bar{\beta}^* -\beta^* \\
\bar{\xi}^* -\xi^* 
\end{bmatrix}
\right\|_2\leq c\epsilon\right\}
\geq &
\mathbb{P} \left\{\left\|
\begin{bmatrix}
y_1(x_1-\bar{x}_1)^\top \\ 
\vdots \\
y_q(x_q-\bar{x}_q)^\top
\end{bmatrix}
\right\|_2\leq \epsilon \right\} \\
\geq&   \mathbb{P}\left\{\sum_{i=1}^q \|n_i\|_2^2\leq \epsilon^2\right\}\\
=&1-\mathbb{P} \left\{
\sum_{i=1}^q \|n_i\|_2^2\geq \epsilon^2 \right\}
\\
\geq 
&1-\frac{1}{\epsilon^2}\mathbb{E} \left\{
\sum_{i=1}^q \|n_i\|_2^2 \right\},
\end{align*}
where the last inequality follows from the Markov's inequality~\citep[Theorem~8.3]{bremaud2001markov}. 
Noting that $\mathbb{E} \{\sum_{i=1}^q \|n_i\|_2^2 \}=q\trace(\int_n nn^\top \gamma(n)\mathrm{d}n)$ while setting $\varepsilon=c\epsilon$ and $\varepsilon_0=c\epsilon_0$ completes the proof.

\section{Proof of Theorem~\ref{tho:optimalnoise_1}}
\label{proof:tho:optimalnoise_1}
Note that
\begin{align*}
\mathcal{L}:=
&\trace(\Pi^{-1}\mathcal{I})+\int_{n}\lambda n^\top n\gamma(n)\mathrm{d}n+\mu\bigg(-1+\int_{n}\gamma(n)\mathrm{d}n \bigg)\\[-.4em]
=&\int_{n}\bigg(\frac{1}{\gamma(n)}\bigg[\frac{\partial \gamma(n)}{\partial n} \bigg]^\top\Pi^{-1}\bigg[\frac{\partial \gamma(n)}{\partial n} \bigg]\\
&\hspace{1in}+(\lambda n^\top n+\mu) \gamma(n)\bigg) \mathrm{d}n-\mu.
\end{align*}
Noting that the cost function and the constraints are convex, the stationary condition of $\mathcal{L}$ (that the variational derivative is equal to zero) is sufficient for optimality. Further, if multiple density functions satisfy the
sufficiency conditions, they all exhibit the same cost. Using Theorem~5.3 in~\citep[p.\,440]{edwards1973advanced}, it can be seen that the extrema must satisfy
\begin{align*}
\lambda n^\top n+\mu&+\frac{1}{\gamma(n)^2}\bigg[\frac{\partial \gamma(n)}{\partial n} \bigg]^\top\Pi^{-1}\bigg[\frac{\partial \gamma(n)}{\partial n} \bigg]\\
&-2\frac{1}{\gamma(n)}\trace ( \Pi^{-1}\mathfrak{D}^2\gamma(n))=0.
\end{align*}
With the change of variable $\gamma(n)=u(n)^2$, the condition can be transformed into
\begin{align*}
\lambda n^\top n+\mu-4\frac{1}{u(n)} \trace(\Pi^{-1}\mathfrak{D}^2u(n))=0.
\end{align*}
It can be seen that the $u(n)=c_0\exp(-n^\top Xn/2)$ with $X=\Pi^{1/2}\lambda^{1/2}/2$ and $c_0=(1/\det(\pi X^{-1}))^{1/4}$ satisfies this equation with $\mu=-4\trace(X)$. 

\section{Proof of Proposition~\ref{prop:Fisher_1}}
\label{app:2}
Define $x=[x_1^\top \; \dots \; x_q^\top]^\top$ and $\bar{x}=[\bar{x}_1^\top \; \dots \; \bar{x}_q^\top]^\top$. Note that the conditional probability density function of $\bar{x}=x+w$ given $x$ is equal to $\tilde{\gamma}(\bar{x}-x)$. Using the Cram\'{e}r-Rao bound~\citep[p.\,169]{cramerraotheorem}, it can be deduced that
\begin{align*}
\mathbb{E}\{\|\Pi(x_i-\hat{x}_i(\bar{x}))&\|_2^2\}\\
\geq \trace\bigg(&\Pi\bigg[\int_{\bar{x} }
\tilde{\gamma}(\bar{x}-x)\bigg[\frac{\partial \log(\tilde{\gamma}(\bar{x}-x))}{\partial x_i} \bigg] \\&\times\bigg[\frac{\partial \log(\tilde{\gamma}(\bar{x}-x))}{\partial x_i} \bigg]^\top\mathrm{d}\bar{x}\bigg]^{-1}\bigg).
\end{align*}
Therefore, 
\begin{align*}
\trace\bigg(&\Pi\bigg[\int_{\bar{x}} 
\tilde{\gamma}(\bar{x}-x)\bigg[\frac{\partial \log(\tilde{\gamma}(\bar{x}-x))}{\partial x_i} \bigg] \\
&\hspace{.4in}\times\bigg[\frac{\partial \log(\tilde{\gamma}(\bar{x}-x))}{\partial x_i} \bigg]^\top\mathrm{d}\revise{\bar{x}}\bigg]^{-1}
\bigg)\\
&\hspace{1in}= \trace(\Pi[(e_i^\top \otimes I_p)\widetilde{\mathcal{I}}(e_i\otimes I_p)]^{-1})
\end{align*}
because
\begin{align*}
\frac{\partial \log(\tilde{\gamma}(\bar{x}-x))}{\partial x_i}
&=(e_i^\top \otimes I_p)\frac{\partial \log(\tilde{\gamma}(w))}{\partial w}\bigg|_{w=\bar{x}-x}.
\end{align*}
Now, by Item~(11) in~\citep[p.\,44]{lutkepohl1997handbook}, it can be seen that
\begin{align*}
\trace(\Pi[(e_i^\top \otimes I_p)&\widetilde{\mathcal{I}}(e_i\otimes I_p)]^{-1})\\
&\geq 1/\trace(\Pi^{-1}(e_i^\top \otimes I_p)\widetilde{\mathcal{I}}(e_i\otimes I_p))\\
&=1/\trace((e_i^\top \otimes I_p)(I_q\otimes \Pi^{-1})\widetilde{\mathcal{I}}(e_i\otimes I_p))\\
&= 1/\trace(((e_ie_i^\top) \otimes I_p)(I_q\otimes \Pi^{-1})\widetilde{\mathcal{I}}).
\end{align*}
Thus,
\begin{align*}
\max_{1\leq i\leq q}1/&\mathbb{E}\{\|x_i-\hat{x}_i(\bar{x})\|_2^2\}\\
&\leq \max_{1\leq i\leq q}\trace(((e_ie_i^\top) \otimes I_p)(I_q\otimes \Pi^{-1})\widetilde{\mathcal{I}})\\
&\leq \sum_{i=1}^q\trace(((e_ie_i^\top) \otimes I_p)(I_q\otimes \Pi^{-1})\widetilde{\mathcal{I}})\\
&=\trace((I_q\otimes \Pi^{-1})\widetilde{\mathcal{I}}).
\end{align*}
The rest follows from \revise{the fact} that $\min_{1\leq i\leq q}\mathbb{E}\{\|x_i-\hat{x}_i(\bar{x})\|_2^2\}=1/(\max_{1\leq i\leq q}1/\mathbb{E}\{\|x_i-\hat{x}_i(\bar{x})\|_2^2\})$.

\section{Proof of Proposition~\ref{proposition:lastonetoremove}}
\label{proof:proposition:lastonetoremove}
Note that $\bar{x}=x+w$ with 
$\bar{x}=[\bar{x}_1^\top \; \dots \; \bar{x}_q^\top]^\top$ and $x=[x_1^\top \; \dots \; x_q^\top]^\top$, the KKT conditions for~\eqref{eqn:optim:1}--depicted in~\eqref{eqn:KKT}--and the KKT conditions for~\eqref{eqn:optim:2} become identical to each other if and only if
$\sum_{i=1}^q (\omega^*)_iy_i(e_i^\top \otimes I_p)w=0$ and $\alpha^{*\top} (e_i^\top \otimes I_p)w=0.$
The rest follows from algebraic manipulation of the equations. 

\section{Proof of Theorem~\ref{tho:noisy_exact}}
\label{proof:tho:noisy_exact}
If $w=\Psi\bar{w}$ with probability one for some random variable $\bar{w}$, it can be seen that
$\Omega w =\Omega\Psi\bar{w}=0.$
Note that $\tilde{\gamma}(w)=0$ for all $w\in\mathrm{im}(\Psi)^\perp$ and $\tilde{\gamma}(w)=\hat{\gamma}(\Psi^\dag w)$ for all $w\in\mathrm{im}(\Psi)$, where $\hat{\gamma}$ is the probability density function of $\bar{w}$. Therefore,
\begin{align*}
\widetilde{\mathcal{I}}
&=\int_{w\in \mathbb{R}^{qp}}
\tilde{\gamma}(w)\bigg[\frac{\partial \log(\tilde{\gamma}(w))}{\partial w} \bigg]\bigg[\frac{\partial \log(\tilde{\gamma}(w))}{\partial w} \bigg]^\top \mathrm{d}w\\[-.4em]
&=\int_{w\in \mathrm{im}(\Psi)}\hspace{-.2in}
\hat{\gamma}(\Psi^\dag w)\bigg[\frac{\partial \log(\hat{\gamma}(\Psi^\dag w))}{\partial w} \bigg]\bigg[\frac{\partial \log(\hat{\gamma}(\Psi^\dag w))}{\partial w} \bigg]^\top \mathrm{d}w\\[-.4em]
&=\int_{\bar{w}}
\hat{\gamma}(\bar{w})\Psi^{\dag\top}\bigg[\frac{\partial \log(\hat{\gamma}(\bar{w}))}{\partial \bar{w}} \bigg]\bigg[\frac{\partial \log(\hat{\gamma}(\bar{w}))}{\partial \bar{w}} \bigg]^\top\Psi^{\dag} \mathrm{d}\bar{w}.
\end{align*}
Further, note that
$\Psi^{\dag}\Psi^{\dag\top}
=\Psi^{\dag}(\Psi^\top)^{\dag}
=(\Psi^\top\Psi)^{\dag}
=I,
$
where the last equality follows from the definition of $\Psi$. Therefore, the optimization problem in~\eqref{eqn:optim:eps_delta} can be translated into
\begin{subequations} \label{eqn:optim:eps_delta_changed}
\begin{align}
\min_{\revise{\hat{\gamma}}} \quad  &\trace\bigg(\Psi^{\dag}(I_q\otimes \Pi^{-1})\Psi^{\dag\top}\nonumber\\
&
\times \int_{\bar{w}}\hat{\gamma}(\bar{w})\bigg[\frac{\partial \log(\hat{\gamma}(\bar{w}))}{\partial \bar{w}} \bigg]\bigg[\frac{\partial \log(\hat{\gamma}(\bar{w}))}{\partial \bar{w}} \bigg]^\top\mathrm{d}\bar{w}\bigg), \\[-.4em]
\st \hspace{.15in} & V_{\bar{w}\bar{w}}\leq mI.
\end{align}
\end{subequations}
Following the same line of reasoning as in the proof of Theorem~\ref{tho:optimalnoise_1}, it can be shown that the solution of~\eqref{eqn:optim:eps_delta_changed} is a Gaussian noise with zero mean and co-variance $V_{\bar{w}\bar{w}}$. Therefore, this problem can be further simplified into
\begin{subequations} 
\begin{align}
\min_{V_{\bar{w}\bar{w}}\geq 0} \quad & \trace\bigg(\Psi^{\dag}(I_q\otimes \Pi^{-1})\Psi^{\dag\top}V_{\bar{w}\bar{w}}^{-1}\bigg), \\[-.4em]
\st \hspace{.23in} & V_{\bar{w}\bar{w}}\leq mI.
\end{align}
\end{subequations}
Because of the inequality constraint in the above optimization problem, $V_{\bar{w}\bar{w}}^{-1}\geq (1/m)I$, and, thus, it can be proved that
\begin{align*}
\trace\bigg(\Psi^{\dag}&(I_q\otimes \Pi^{-1})\Psi^{\dag\top}V_{\bar{w}\bar{w}}^{-1}\bigg)\\
&=\trace\bigg((I_q\otimes \Pi^{-1})^{1/2}\Psi^{\dag\top}V_{\bar{w}\bar{w}}^{-1}\Psi^{\dag}(I_q\otimes \Pi^{-1})^{1/2}\bigg)\\[-.4em]
&\geq (1/m)\trace\bigg((I_q\otimes \Pi^{-1})^{1/2}\Psi^{\dag\top}\Psi^{\dag}(I_q\otimes \Pi^{-1})^{1/2}\bigg)\\[-.4em]
&=(1/m)\trace\bigg(\Psi^{\dag}(I_q\otimes \Pi^{-1})\Psi^{\dag\top}\bigg).
\end{align*}
The lower bound on the cost function can be clearly achieved by selecting $V_{\bar{w}\bar{w}}=mI$.

\section{Proof of Proposition~\ref{prop:lowerbound_general}}
\label{proof:prop:lowerbound_general}
The optimality condition is$\nabla_\phi \ell(\varphi;\{(x_i,y_i)\}_{i=1}^q)=0.$ Small variations of the dataset shows that
$$\mathfrak{D}_\varphi^2 \ell(\varphi;\{(x_i,y_i)\}_{i=1}^q)\nabla_{x_j}\varphi+\nabla_{x_j}\nabla_\phi \ell(\varphi;\{(x_i,y_i)\}_{i=1}^q)=0.$$ Therefore, if $\mathfrak{D}_\varphi^2 \ell(\varphi;\{(x_i,y_i)\}_{i=1}^q)$ is invertible, it can be shown that 
\begin{align*}
\nabla_{x_j}\varphi=&\hspace{-.03in}-\hspace{-.03in}[\mathfrak{D}_\varphi^2 \ell(\varphi;\{(x_i,y_i)\}_{i=1}^q)]^{-1}\hspace{-.02in}\nabla_{x_j}\hspace{-.03in}\nabla_\phi\hspace{-.02in} \ell(\varphi;\{(x_i,y_i)\}_{i=1}^q).
\end{align*}
The Taylor's Theorem shows that
\begin{align*}
\|\bar{\varphi}-\varphi\|_2\leq \sum_{j=1}^q\bigg[&\|[\mathfrak{D}_{\varphi\varphi}^2 \ell(\varphi';\{(x_i,y_i)\}_{i=1}^q)]^{-1}\\
&\times\mathfrak{D}^2_{x_i\varphi} \ell(\varphi';\{(x_i,y_i)\}_{i=1}^q)\|_2\|x_i-\bar{x}_i\|_2\bigg],
\end{align*}
 for some $\varphi'=s\varphi+(1-s)\bar{\varphi}$ and $s\in[0,1]$. Therefore, there exist $c\geq 1$ and $\epsilon_0>0$ (due to continuity of the second order derivatives) such that if $\sum_{i=1}^q\|x_i-\bar{x}_i\|_2 \leq \epsilon<\epsilon_0$, then $\|\bar{\varphi} -\varphi\|_2\leq c\epsilon.$ The Markov's inequality~\citep[Theorem~8.3]{bremaud2001markov} implies that
\begin{align*}
\mathbb{P}\left\{\sum_{i=1}^q\|x_i-\bar{x}_i\|_2 \leq \epsilon/c\right\}
&=\mathbb{P}\left\{\sum_{i=1}^q\|n_i\|_2 \leq \epsilon/c\right\}\\
&=\mathbb{P}\left\{\bigg(\sum_{i=1}^q\|n_i\|_2\bigg)^2 \leq \epsilon^2/c\right\}\\
&\geq 1-\frac{c^2}{\epsilon^2}\mathbb{E}\left\{\bigg(\sum_{i=1}^q\|n_i\|_2\bigg)^2\right\}.
\end{align*}
Furthermore, 
\begin{align*}
\mathbb{E}\hspace{-.03in}\left\{\hspace{-.03in}\bigg(\sum_{i=1}^q\|n_i\|_2\hspace{-.03in}\bigg)^{\hspace{-.03in}2}\hspace{-.03in}\right\}
\hspace{-.03in}&=\hspace{-.03in}\mathbb{E}\left\{\sum_{i=1}^qn_i^\top n_i\hspace{-.03in}+\hspace{-.03in}\sum_{i=1}^q\sum_{j\neq i}\sqrt{n_i^\top n_i n_j^\top n_j} \right\}\\
&\leq q\trace(V_{nn})\hspace{-.03in}+\hspace{-.03in}\sum_{i=1}^q\hspace{-.03in}\sum_{j\neq i}\hspace{-.03in}\sqrt{\mathbb{E}\left\{n_i^\top n_i n_j^\top n_j\right\} }\\
&=q\trace(V_{nn})+\sum_{i=1}^q\sum_{j\neq i}\sqrt{\trace(V_{nn})^2 }\\
&=q^2\trace(V_{nn}).
\end{align*}
where the inequality follows from the Jensen's inequality (and the concavity of the square root function). 

\end{document}